\documentclass[a4paper]{article}
\usepackage{Odyssey2018}
\usepackage{epsfig,amssymb,amsmath,multirow,boldline,graphicx,makecell}
\ninept
\usepackage{kotex}
\usepackage{color,soul}

\title{Convolutional Neural Networks and Language Embeddings\\for End-to-End Dialect Recognition }

\makeatletter
\def\name#1{\gdef\@name{#1\\}}
\makeatother
\name{{\em Suwon Shon$^1$, Ahmed Ali$^2$, James Glass$^1$}
}

\address{Massachusetts Institute of Technology, Cambridge, MA, USA$^1$  \\
Qatar Computing Research Institute, HBKU, Doha, Qatar$^2$\\
{\small \tt \{swshon,glass\}@mit.edu \qquad amali@hbku.edu.qa} }
\begin{document}
\maketitle

\begin{abstract}
Dialect identification (DID) is a special case of general language identification (LID), but a more challenging problem due to the linguistic similarity between dialects. In this paper, we propose an end-to-end DID system and a Siamese neural network to extract language embeddings. We use both acoustic and linguistic features for the DID task on the Arabic dialectal speech dataset: Multi-Genre Broadcast 3 (MGB-3). The end-to-end DID system was trained using three kinds of acoustic features: Mel-Frequency Cepstral Coefficients (MFCCs), log Mel-scale Filter Bank energies (FBANK) and spectrogram energies. We also investigated a dataset augmentation approach to achieve robust performance with limited data resources.  Our linguistic feature research focused on learning similarities and dissimilarities between dialects using the Siamese network, so that we can reduce feature dimensionality as well as improve DID performance. The best system using a single feature set achieves 73\% accuracy, while a fusion system using multiple features yields 78\% on the MGB-3 dialect test set consisting of 5 dialects. The experimental results indicate that FBANK features achieve slightly better results than MFCCs. Dataset augmentation via speed perturbation appears to add significant robustness to the system.  Although the Siamese network with language embeddings did not achieve as good a result as the end-to-end DID system, the two approaches had good synergy when combined together in a fused system.

\end{abstract}

\section{Introduction}
A significant step forward in speaker and language identification (LID) was obtained by combining i-vectors and Deep Neural Networks (DNNs)~\cite{Cardinal2015,Richardson2015,Dehak2011b}. The task of dialect identification (DID) is relatively unexplored compared to speaker and language recognition.  One of the main reasons is due to a lack of common datasets, while another reason is that DID is often regarded as a special case of LID, so researchers tend to concentrate on the more general problems of language and speaker recognition. However, DID can be an extremely challenging task since the similarities among language dialects tends to be much higher than those used in the more general LID task.

Arabic is an appropriate language for which to explore DID, due to its uniqueness and widespread use. While 22 countries in the Arab world use Modern Standard Arabic (MSA) as their official language, citizens use their own local dialect in their everyday life.  Arabic dialects are historically related and share Arabic characters, however, they are not mutually comprehensible. 
Arabic DID therefore poses different challenges compared to other language dialects containing comprehensible vernacular. Arabic dialects are challenging to distinguish because they belong to the same language family.  Arabic dialects typically share a common phonetic inventory and other linguistic feature like characters, so words and phonemes can be utilized by automatic speech recognition (ASR) contrary to general LID.

The natural language processing (NLP) community has tended to partition the Arabic language into 5 broad categories: Egyptian (EGY), Levantine (LEV), Gulf (GLF), North African (NOR), and MSA.  The Multi Genre Broadcast(MGB) challenge\footnote{www.mgb-challenge.org} committee established an Arabic dialect dataset in 2016, and holds a DID challenge series that attracts interest in Arabic DID from researchers in both the NLP and speech communities. The MGB-3 dataset contains 5 dialects with 63.6 hours of training data and 10.1 hours of test data. A DID benchmark of the task was attained using the i-vector framework using bottleneck features~\cite{Ali2017}.  Using linguistic features such as words and characters results in similar performance to that obtained via acoustic features with a Convolutional Neural Network (CNN)-based backend~\cite{Khurana2017,Ali2016,Maryam2018}. Since the linguistic feature space is different from the acoustic feature space, a fusion of the results from both feature representations has been shown to be beneficial~\cite{Ali2016,Khurana2017,Najafian2016}. A recent study achieves 70\%~\cite{Shon2017c} for a single system and 80\%~\cite{Bulut2017} for a fused system. Given that there are only 5 dialects, these accuracies are relatively low, affirming the difficulty of this DID task.

Recently, many Deep Neural Network (DNN)-based end-to-end and speaker embedding approaches have achieved impressive results for text independent speaker recognition and LID~\cite{Jin2017, Trong2016, Nagraniy2017, DavidSnyder2016}. This research indicates that these methods perform close to or slightly better than traditional i-vectors, though it has focused on improving upon i-vectors, so there is room for further detailed analyses of various task conditions.

In this work, we propose two different approaches for DID using both acoustic and linguistic features. For acoustic features, we explore an end-to-end DID model based on a CNN with global average pooling. We examine three representations based on Mel-Frequency Cepstral Coefficients (MFCCs), log Mel-scale filter banks energies (FBANK) and spectrogram energies.  We employ data augmentation via speech and volume perturbation to analyze the impact of dataset size.  Finally, we analyze the effectiveness of training with random utterance segmentation, as a strategy to cope with short input utterances. 

For language embeddings, we adopt a Siamese neural network~\cite{Bromley1993} model using a cosine similarity metric to learn a dialect embedding space based on text-based linguistic features.  The network is trained to learn similarities between the same dialects and dissimilarities between different dialects. Linguistic feature are extracted from an automatic speech recognizer (ASR), and based on words, characters and phonemes.  Siamese neural networks are usually applied on verification tasks for end-to-end systems, however a previous study showed that this approach could be also applied on an identification task to make the original features more robust~\cite{Shon2017c}.  We applied this approach on linguistic features to extract language embeddings to improve DID performance while reducing the enormous feature dimensionality.  Python and Tensorflow code for the end-to-end system\footnote{https://github.com/swshon/dialectID\_e2e} and Siamese network language embeddings\footnote{https://github.com/swshon/dialectID\_siam} is available, to allow others to reproduce our results, and apply these ideas to similar LID and speaker verification tasks.

\section{Dialect Identification Baseline}

\subsection{Acoustic features}
For LID and DID, the i-vector is regarded as the state-of-the-art for general tasks, while recent studies show combinations with DNN-based approaches produce competitive performance~\cite{Richardson2015}.  Apart from the basic i-vector approach, bottleneck (BN) features extracted from an ASR acoustic model were successfully applied to LID~\cite{Cardinal2015,Matejka2014,Richardson2015,Khurana2017, Ali2016}.  Although the acoustic model is usually trained on single language, it performs reasonably well on LID tasks \cite{Matejka2014}.  The mismatch does appear to matter on speaker recognition tasks however~\cite{Shon2017a,Shon2017b}. A stacked BN feature can be extracted from a second DNN that has a BN layer, and its input is the output of the first BN layer with context expansion~\cite{Cardinal2015,Matejka2014}. A Gaussian Mixture Model - Universal Background Model (GMM-UBM) and a total variability matrix are trained in an unsupervised manner with BN features from a large dataset. Although the main purpose of LID is to obtain phonetic characteristics of each languages while suppressing individual speaker's voice characteristics, the training scheme is exactly the same as training a GMM-UBM and total variability matrix for speaker recognition. The only difference is that the subsequent projection of the i-vector is based on measuring the distances with other languages as opposed to other speakers.

\begin{figure}[t]
\centerline{\includegraphics[width=8cm]{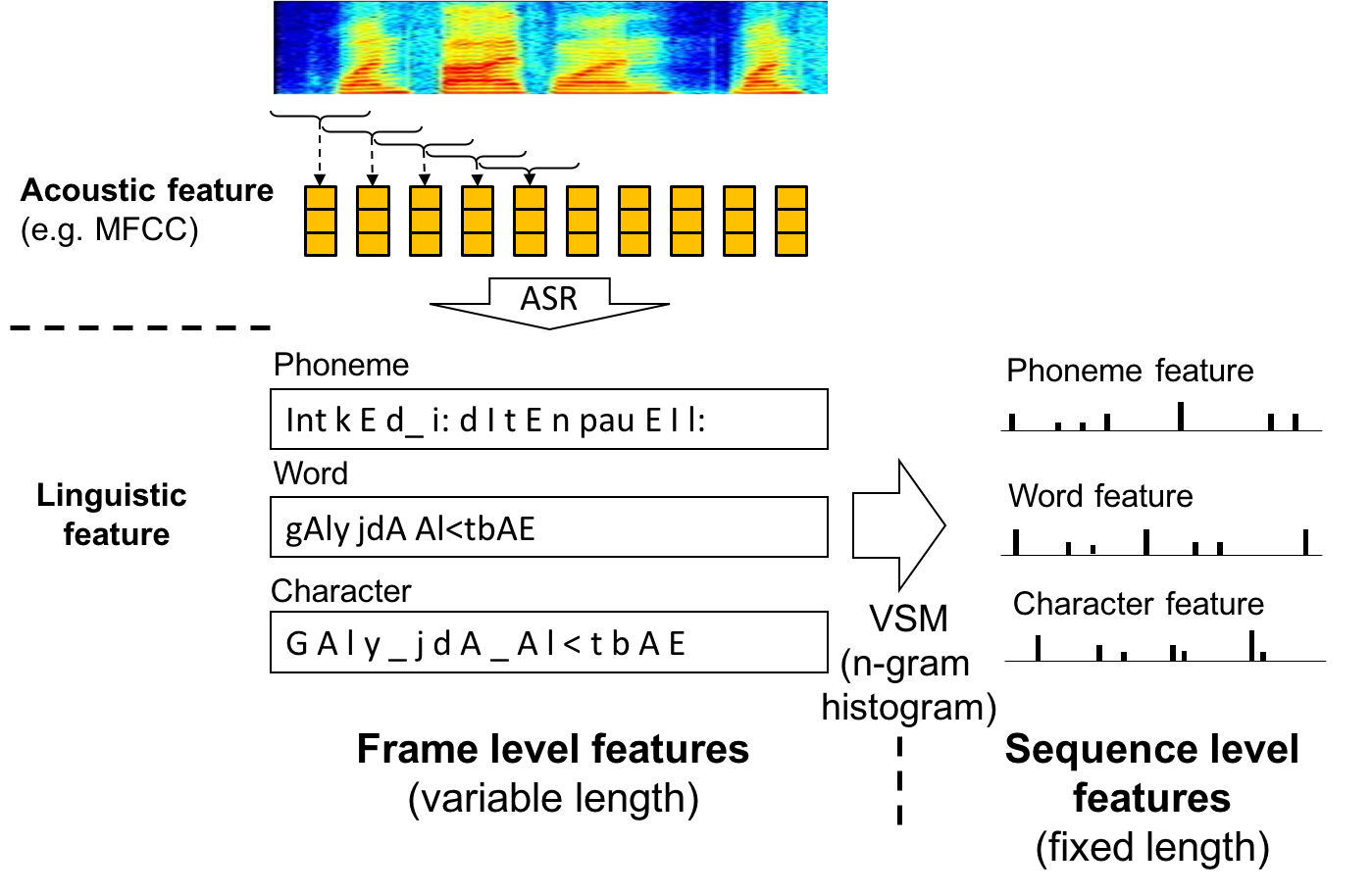}}
\caption{Illustration of acoustic and linguistic feature creation.}
\label{fig:features}
\end{figure}

\subsection{Linguistic features}

While the inherent similarity between dialects makes DID a more difficult task than LID, DID can take advantage of linguistic features because dialects tend to share a common phonetic inventory. Word and character sequences can be extracted using a state-of-the-art ASR system, such as the Time delayed Neural Network (TDNN)~\cite{Peddinti2015} or Bidirectional Long Short-Term Memory Recurrent Neural Network (BLSTM-RNN) acoustic model with RNN-based language model rescoring. Extracted text sequence can then be converted into a Vector Space Model(VSM)~\cite{Ali2016}. The VSM represents each utterance as a fixed length, sequence level, high-dimensional sparse vector $\vec{u}$:
\begin{equation}
\vec{u} = (f(u,x_1), f(u,x_2), ... , f(u,x_D))
\label{eq1}
\end{equation}
where $f(u,x_i)$ is the number of times word $x_i$ occurs in utterance $u$, and $D$ is the dictionary size. The vector $\vec{u}$ can also consist of an $n$-gram histogram. Figure~\ref{fig:features} illustrates the linguistic feature extraction concept based on acoustic features.

For the MGB-3 Arabic dataset, the tri-gram word dictionary size $D$ is 466k, and most of $f(u,x_i)$ is 0. Due to this high dimensionality, the kernel trick method used with Support Vector Machine's (SVMs) is a suitable approach for classification. Another linguistic feature, the character and phoneme feature, can also be extracted by an ASR system or stand-alone phonetic recognizer. Although both ASR and phonetic recognizers start from an acoustic input, the linguistic feature space consists of higher level features that contain semantic information of variable duration speech segments.  While an ASR system, and an acoustic feature-based speaker recognition or LID/DID system are vulnerable to domain mismatched conditions~\cite{Aronowitz2014,Garcia-Romero2014,Shum2014,Shon2017}, the affected results in the higher-level linguistic spaces will be different. Thus, linguistic feature-based approaches could complement acoustic feature-based approaches to produce a superior fusion result.

\section{End-to-End DID with Acoustic Features}

Recently, end-to-end approaches which do not use i-vector techniques showed impressive results on both LID and speaker recognition~\cite{Jin2017, Trong2016, Nagraniy2017, DavidSnyder2016}. They did not start from raw waveforms, but from acoustic features ranging from MFCCs to spectrograms. The deep learning models consisted of CNNs, RNNs and Fully Connected (FC) DNN layers. Recent studies~\cite{DavidSnyder2016, Nagraniy2017} report combinations of CNN and FC with a global pooling layer obtains the best result for a speaker representation from text-independent and variable length speech inputs. The global pooling layer has a simple function such as averaging sequential inputs, but it effectively convert frame level representations to utterance level representations, which is advantageous for speaker recognition and LID.
The size of the final softmax layer is determined by the task specific speaker or language labels. For identification tasks, the Softmax output can be used directly as a score for each class. For verification tasks, usually the activation of the last hidden layer is extracted as a speaker or language representation vector, and is used to calculate a distance between enrollment and test data.

\begin{figure}[t]
\centerline{\includegraphics[width=5cm]{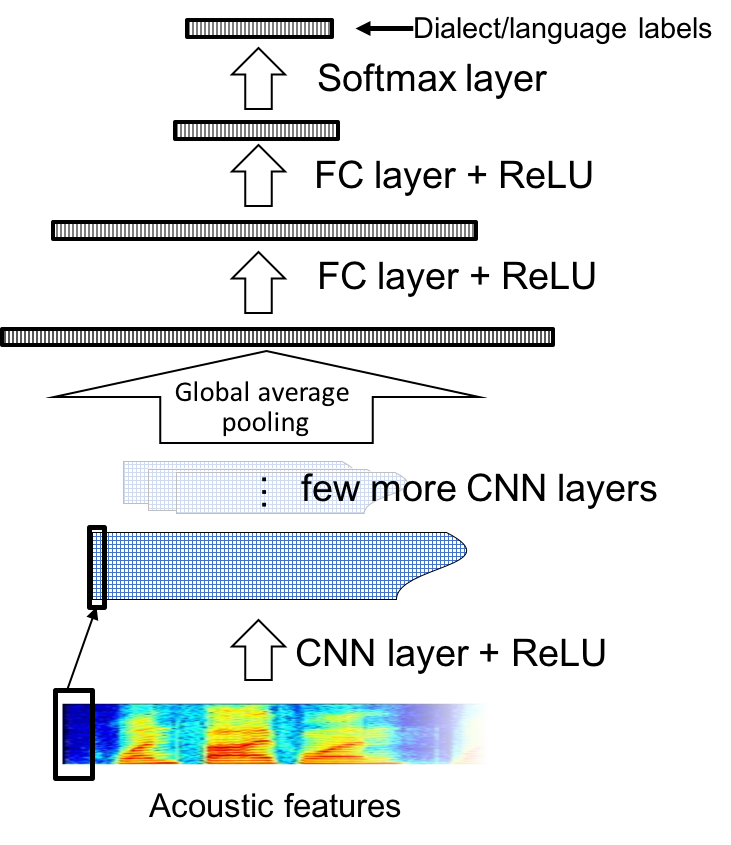}}
\caption{End-to-End dialect identification system structure.}
\label{fig:e2e_network}
\end{figure}

Our end-to-end system is based on~\cite{DavidSnyder2016, Nagraniy2017}, but instead of a VGG~\cite{Simonyan2014} or Time Delayed Neural Network, we used four 1-dimensional CNN (1d-CNN) layers (40$\times$5 - 500$\times$7 - 500$\times$1 - 500$\times$1 filter sizes with 1-2-1-1 strides and the number of filters is 500-500-500-3000) and two FC layers (1500-600) that are connected with a Global average pooling layer which averages the CNN outputs to produce a fixed output size of 3000$\times$1. After global average pooling, the fixed length output is fed into two FC layers and a Softmax layer. The structure of end-to-end dialect identification system is represented in Figure \ref{fig:e2e_network}. The four CNN layers span a total of 11 frames, with strides of 2 frames over the input. For Arabic DID, the Softmax layer size is 5. 
We examined the FBANK, MFCC and spectrogram as acoustic features. Filter size on first CNN layer changes to 200$\times$5 when we use Spectrogram. The Softmax output was used to compute similarity scores of each dialect for identification.

\subsection{Internal Dataset augmentation}

Neural network-based deep learning models are but the most recent example of the age-old ASR adage that ``there's no data like more data."  These models are able to absorb large quantities of training data, and generally do better, the more data they are exposed to.  Given a fixed training data size, researchers have found that augmenting the corpus size by artificial means can also be effective.  In this paper, we explored internal dataset augmentation methods.  One augmentation approach is segmentation of the training dataset into small chunks~\cite{Nagraniy2017, DavidSnyder2016, Jin2017}.  We did not segment the dataset in advance, but did random segmentations on mini-batches. The random length was selected to have a uniform distribution between 2 and 10 seconds with 1 second intervals including original length. From previous studies, it was found that random segmentation yields better performance on short utterances~\cite{Nagraniy2017, DavidSnyder2016, Jin2017}. However, the investigations did not show the performance before applying this method, so it is difficult to judge how much gains were achieved from this approach. We analyzed performances before and after applying random segmentation on various test set lengths.

The other technique for augmentation is perturbation of the original dataset in term of speed~\cite{Ko2015} which is used ASR. The idea is to perturb the audio to play slightly faster or slower. We investigated the effectiveness of speed and volume perturbation on the DID task using speed factors of 0.9 and 1.1 and volume factors of 0.25 and 2.0.

\begin{figure}[t]
\centerline{\includegraphics[width=7.5cm]{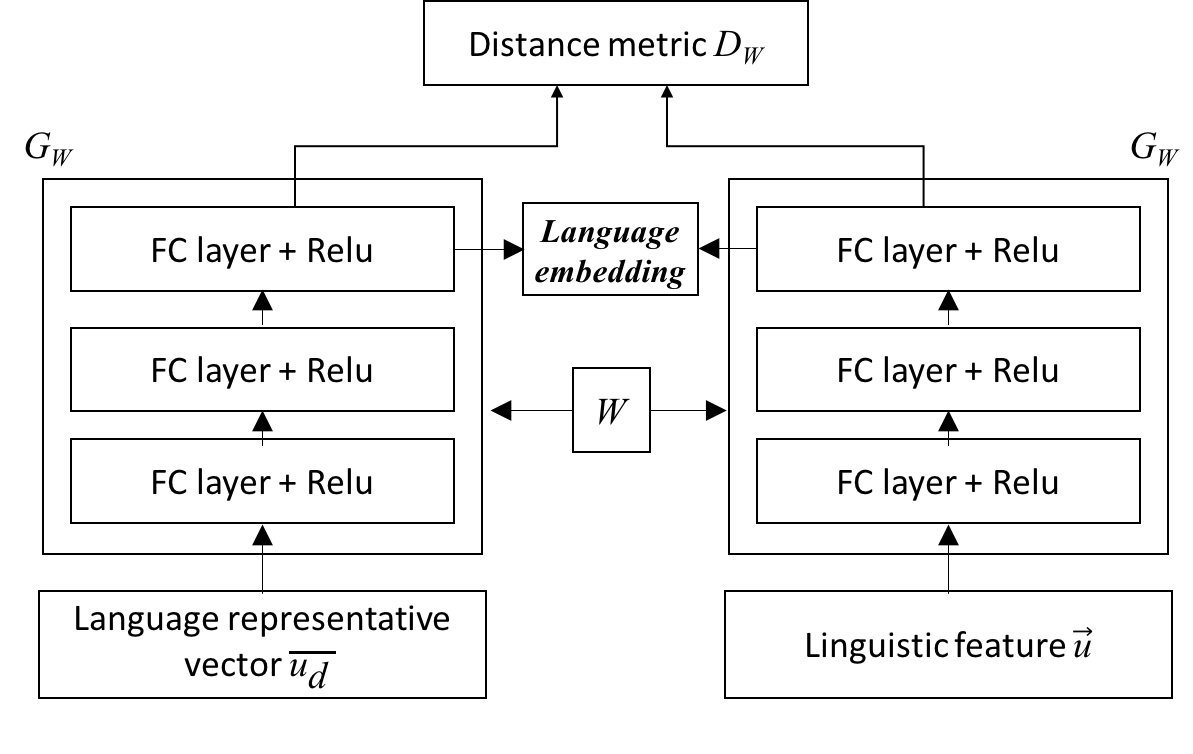}}
\caption{Siamese neural network structure.}
\label{fig:siam_network}
\end{figure}

\section{DID Linguistic Feature Embeddings}

In previous study~\cite{Khurana2017}, they explored CNN back-end for the linguistic feature using cross-entropy objective function with softmax output of the network. For identification task, it is natural to consider that the cross-entropy objective function between class labels and output of the network. However, our recent study~\cite{Shon2017c} shows that using the Euclidean distance loss function between the label and the cosine distance of the NN output pair which is usually adopted on a binary classification task is still useful for identification task by learning similarity and dissimilarity between classes. Thus, we adopted a Siamese neural network approach to extract language embeddings from linguistic features. The Siamese neural network has two parallel neural networks, $G_W$, which shares the same set of weights and biases $W$ as shown in Figure~\ref{fig:siam_network}. While the network has a symmetric structure, the input pair to the network consists of language representative vectors $\overline{u_d}$ and linguistic feature $\vec{u}$, where $d$ is language/dialect index and $i$ is utterance index. The language representative vector can be represented as $\overline{u_d}=(1/n_d)\sum_{i=1}^{n_d}\vec{u}^d_i$, where $n_d$ is the number of utterances for each dialect $d$ and $\vec{u}^d_i$ is a segment level feature of dialect $d$. Thus, the pair of network inputs is the combination of the language representative vectors and all linguistic features.
Let $Y$ be the label for the pair, where $Y=1$ if the pair belongs to the same language/dialect, and $Y=-1$ if a different language/dialect. To optimize the network, we use a Euclidean distance loss function with stochastic gradient descent (SGD) between the label and the cosine similarity of the pair, $D_W$, where
\begin{equation}
L(\overline{u_d},\vec{u},Y) = || Y - D_W(\overline{u_d},\vec{u}) ||_2^2 
\end{equation}
FC layers have 1500-600-200 neurons. All layers use Rectified Linear Unit (ReLU) activations.
From the network $G_W$, language embeddings can be extracted from both language representative vectors and linguistic features by obtaining activations from the last hidden layer (200 dimensions).

\section{Dialectal Speech Dataset}
The MGB-3 dataset partitions are shown in Table~\ref{tab:data}. Each partition consists of fives Arabic dialects : EGY, LEV, GLF, NOR, MSA.  Detailed corpus statistics are found in~\cite{Ali2017}. Although the development set is relatively small compared to the training set,  it matches the test set channel conditions, and thus provides valuable information about the test domain.

\begin{table}[ht]
\centering
\resizebox{0.9\linewidth}{!}{%
\begin{tabular}{cV{2}c|c|c}
\hlineB{2}
\begin{tabular}[c]{@{}c@{}}Dataset\end{tabular}& \begin{tabular}[c]{@{}c@{}}Training\end{tabular}& \begin{tabular}[c]{@{}c@{}}Development\end{tabular}   & \begin{tabular}[c]{@{}c@{}}Test\end{tabular}  \\ \hlineB{2}
Utterances & 13,825 & 1,524 & 1,492\\  \hline
Size & 53.6 hrs & 10 hrs & 10.1 hrs\\  \hline
\begin{tabular}[c]{@{}c@{}}Channel\\ (recording)\end{tabular}& \begin{tabular}[c]{@{}c@{}}Carried out\\ at 16kHz\end{tabular} & \multicolumn{2}{c}{\begin{tabular}[c]{@{}c@{}}Downloaded directly from\\ a high-quality video server\end{tabular}} \\ \hlineB{2}
\end{tabular}%
}
\caption{MGB-3 Dialectal Arabic Speech Dataset Properties.}
\label{tab:data}
\end{table}

\begin{table*}[t]
\centering
\resizebox{0.8\textwidth}{!}{%
\begin{tabular}{c|ccc|ccc}
\hline
\multirow{2}{*}{Feature} & \multicolumn{3}{c|}{Maximum} & \multicolumn{3}{c}{Converged} \\ \cline{2-7} 
 & Accuracy(\%) & EER(\%) & C\textsubscript{avg} & Accuracy(\%) & EER(\%) & C\textsubscript{avg} \\ \hline
MFCC & 65.55 & 20.24 & 19.92 & 61.33 & 21.95 & 21.53 \\
FBANK & 64.81 & 20.22 & 19.91 & 61.26 & 22.12 & 21.79 \\
Spectrogram & 57.57 & 24.48 & 24.49 & 54.22 & 25.90 & 25.09 \\ \hline
\end{tabular}%
}
\caption{Performance on end-to-end dialect identification system by features}
\label{tab:perform_e2e}
\end{table*}

\section{DID Experiments}

\subsection{Implementation and Training}

For i-vector implementation, the BN features were extracted from an ASR system trained for MSA speech recognition~\cite{Khurana2017a} with a similar configuration as \cite{Cardinal2015}. A DNN that extracted BN features was trained on 60 hours of manually transcribed Al-Jazeera MSA news recordings~\cite{Ali2014}. Then, the GMM-UBM with 2,048 mixtures and a 400 dimensional i-vector extractor was trained based on the BN features. The detailed configuration is described in \cite{Ali2016}. After extracting i-vectors, a whitening transformation and length normalization are applied. Similarity scores were calculated by Cosine Similarity (CS) and Gaussian Backend (GB). Linear Discriminant Analysis (LDA) was applied after extracting i-vectors with language labels. On this experiment, LDA reduces the 400 dimensional i-vector to 4 dimensions.

\begin{figure}[t]
\centerline{\includegraphics[width=8cm]{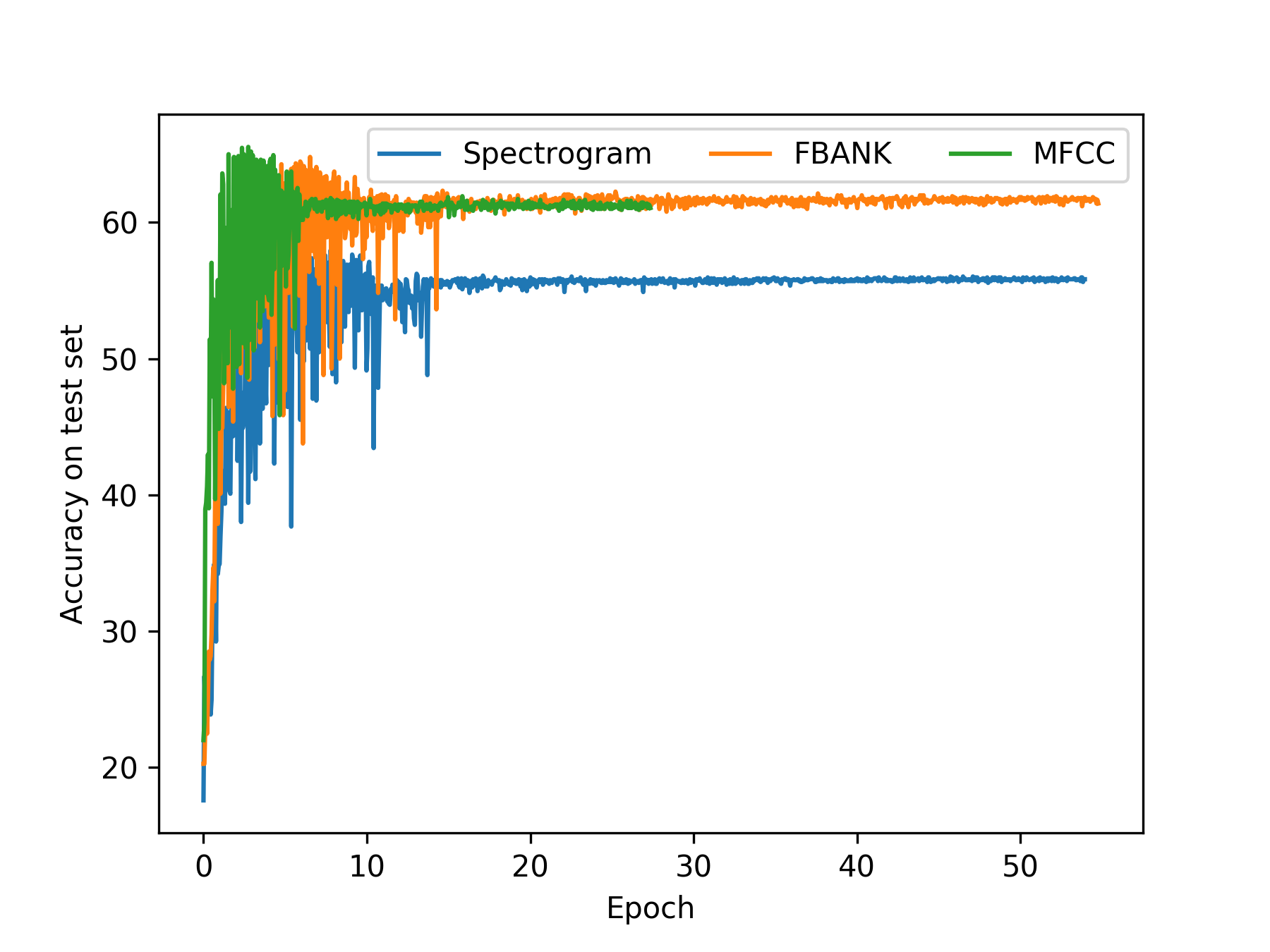}}
\caption{End-to-end DID accuracy by epoch.}
\label{fig:accuracy_feat}
\end{figure}

\begin{figure}[t]
\centerline{\includegraphics[width=8cm]{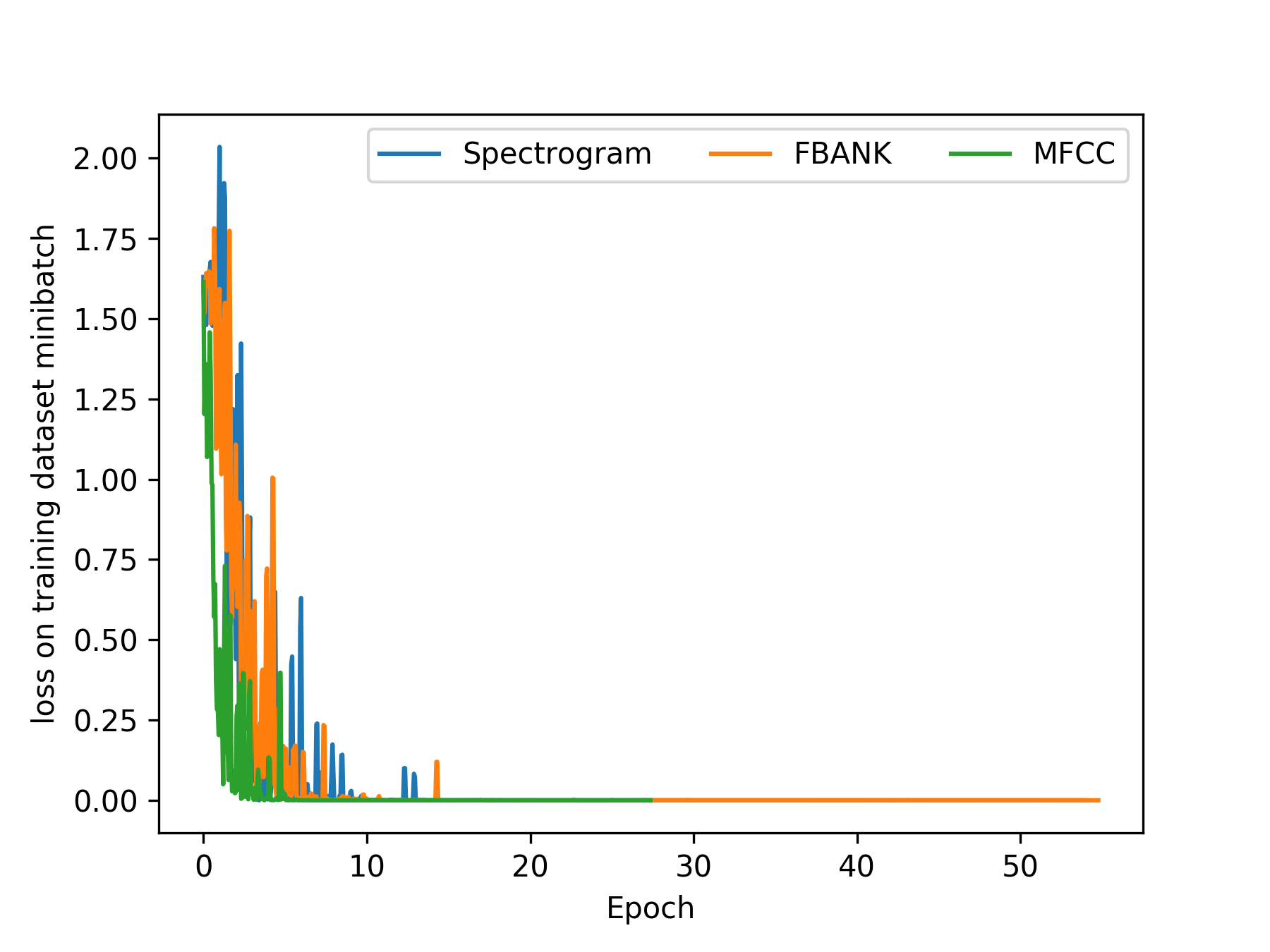}}
\caption{End-to-end DID mini-batch training loss by epoch.}
\label{fig:loss_feat}
\end{figure}

\begin{figure}[t]
\centerline{\includegraphics[width=8cm]{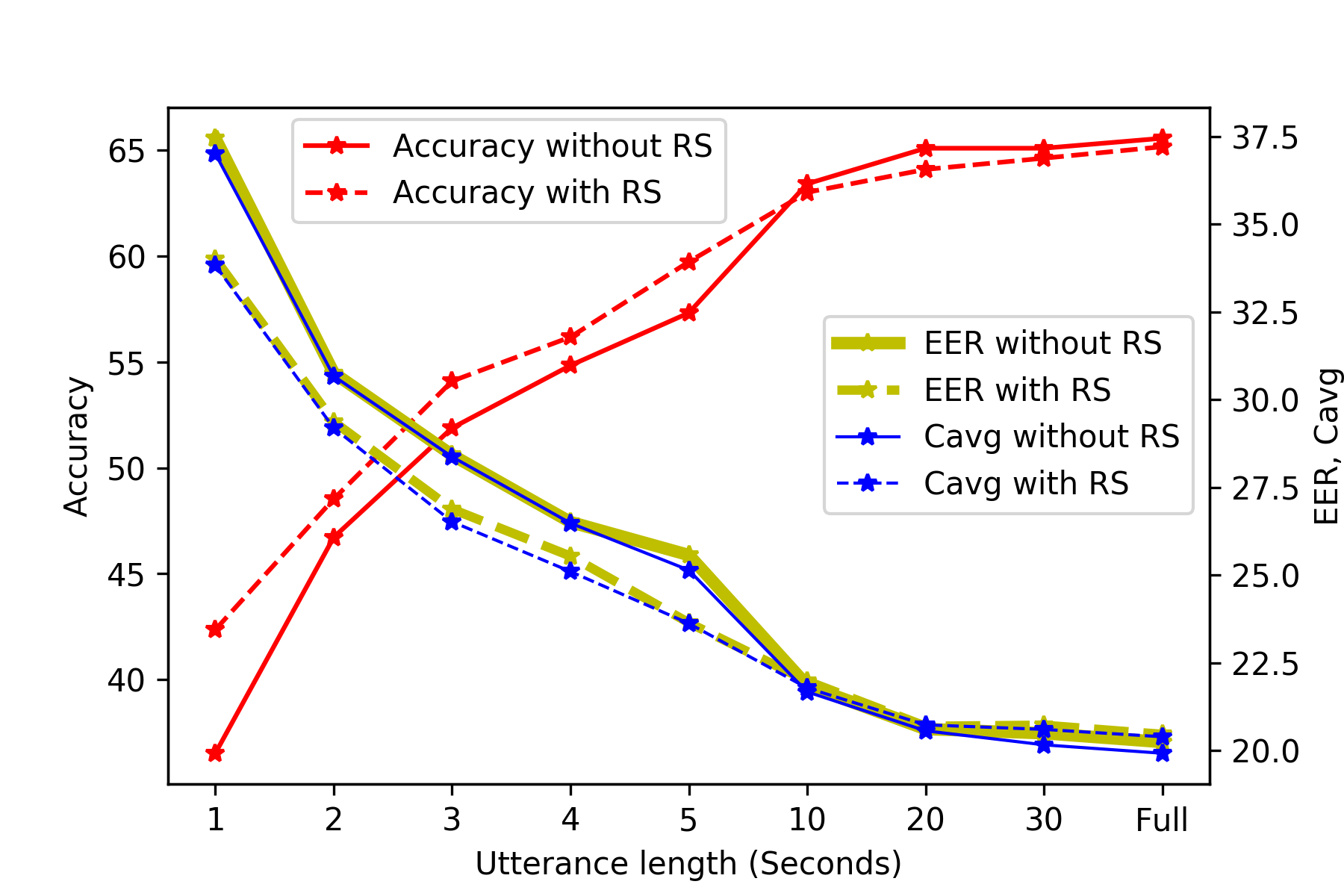}}
\caption{Performance comparison with and without Random Segmentation (RS) on training dataset.}
\label{fig:random_segment}
\end{figure}

For phoneme features, we used four different phoneme recognizers: Czech (CZ), Hungarian (HU) and Russian (RU) using narrow-band models, and English (EN) using a broadband model~\cite{Schwarza}. For words and character features, word sequences are extracted using an Arabic ASR system built as part of the MGB-2 challenge on Arabic broadcast speech~\cite{Ali2016a}. From the word sequence, character sequences can also be obtained by splitting words into characters. Finally, phoneme, character and word feature can be represented in VSM by using each sequence. We use unigrams for word features, and trigrams for character and phoneme feature. The word and character feature dimensions are 41,657 and 13,285, respectively. Phoneme feature dimensions are 47,466 (CZ), 50,320 (HU), 51,102 (RU) and 33,660 (EN). An SVM was used to measure similarity between the test utterance and 5 dialects~\cite{Ali2016}.

To train the end-to-end dialect identification system, we used 3 features: MFCC, FBANK and spectrogram. The structure of the DNN is shown in Figure~\ref{fig:e2e_network} and described fully in Section 3. The SGD learning rate was 0.001 with decay in every 50000 mini-batches with a factor of 0.98. ReLUs were used for activation nonlinearities.  We used both the training and development dataset for training the DNN and excluded 10\% of development dataset of each dialect for a validation set. For acoustic input, the FFT window length was 400 samples with a 160 sample hop which is equivalent to 25ms and 10ms respectively for 16kHz audio. A total of 40 coefficients were extracted for MFCCs and FBANKs, and 200 for spectrograms. All features were normalized to have zero-mean and unit variance.

To train the Siamese network, we made utterance pairs from the training and development sets. Since the number of true ("same dialect") pairs is naturally lower than false ("different dialect") pairs, we adjust the training batch to have the same 50\%:50\% ratio of true and false pairs. Since the test set domain is mismatched with the training set, the development dataset is very important because it contains channel information of the target domain. Thus, we expose development set pairs more than training set pairs by a factor of 5.

\begin{figure}[t]
\centerline{\includegraphics[width=6cm]{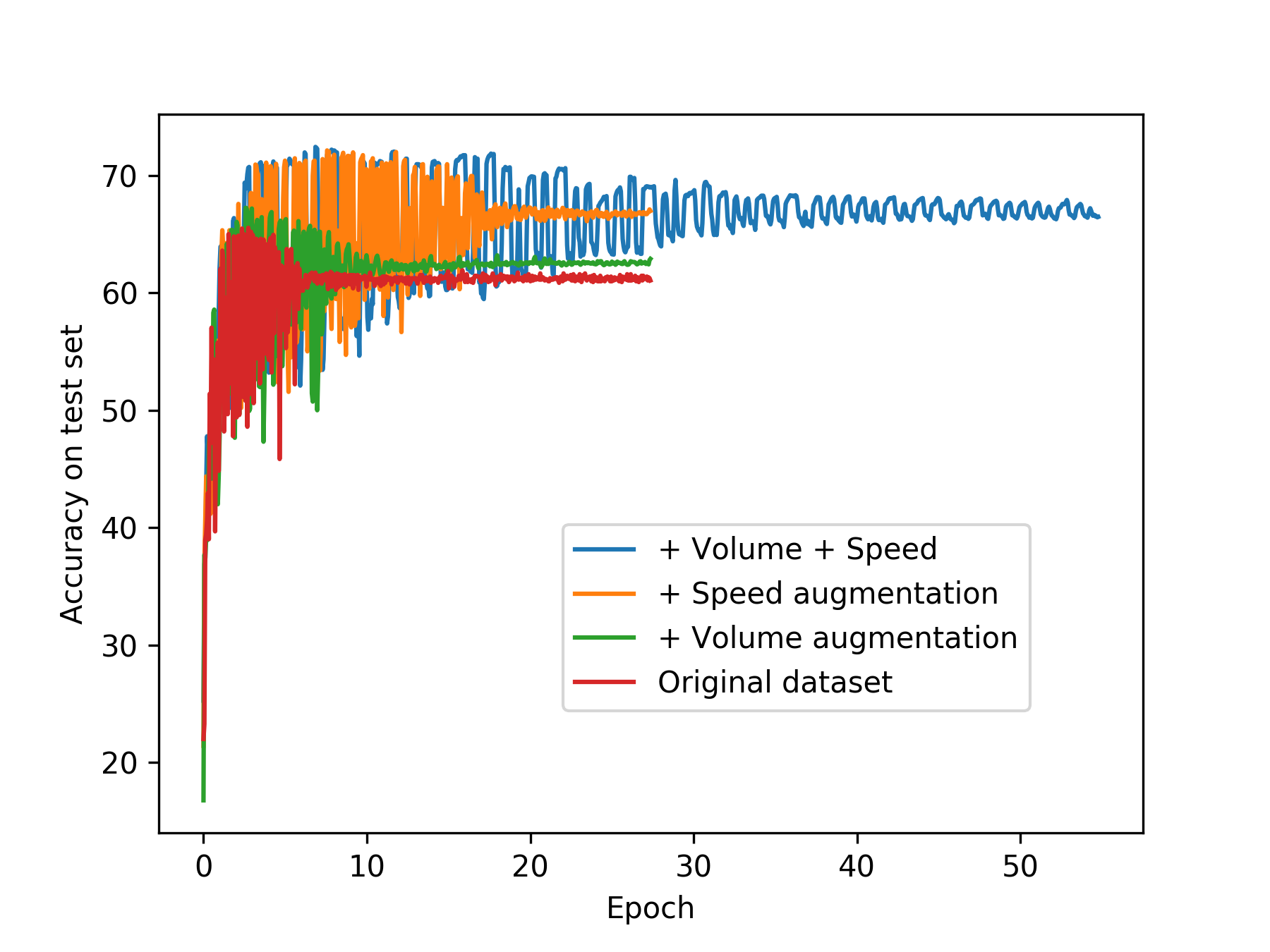}}
\caption{DID accuracy by epoch using augmented data.}
\label{fig:accuracy_aug}
\end{figure}

\begin{table*}[t]
\centering
\resizebox{0.8\textwidth}{!}{%
\begin{tabular}{c|ccc|ccc}
\hline
\multirow{2}{*}{\begin{tabular}[c]{@{}c@{}}Augmentation method\\ (feature = MFCC)\end{tabular}} & \multicolumn{3}{c|}{Maximum} & \multicolumn{3}{c}{Converged} \\ \cline{2-7} 
 & Accuracy(\%) & EER(\%) & C\textsubscript{avg} & Accuracy(\%) & EER(\%) & C\textsubscript{avg} \\ \hline
Volume & 67.49 & 20.37 & 20.00 & 62.47 & 21.55 & 21.08 \\
Speed & 70.51 & 17.54 & 17.39 & 65.42 & 19.87 & 19.19 \\
Volume and speed & 70.91 & 17.79 & 17.93 & 67.02 & 19.37 & 19.01 \\ \hline
\end{tabular}%
}
\caption{Performance on end-to-end DID system using augmented dataset.}
\label{tab:perform_e2e_aug}
\end{table*}

\begin{table}[t]
\centering
\setlength\doublerulesep{0.5pt}
\begin{tabular}{l|ccc}
\hlineB{2}
\multicolumn{1}{c|}{\begin{tabular}[c]{@{}c@{}}Feature\\(on augmented dataset)\end{tabular}} & Accuracy(\%) & EER(\%) & C\textsubscript{avg} \\ \hlineB{2}
MFCC & 70.91 & 17.79 & 17.93 \\
FBANK      & 71.92 & 18.01 & 17.63 \\
Spectrogram     & 68.83 & 18.70 & 18.69 \\ \hlineB{2}
\end{tabular}
\caption{Performance on end-to-end DID system using augmented dataset.}
\label{tab:perform_e2e_aug_byfeat}
\end{table}

\begin{table}[b]
\centering
\setlength\doublerulesep{0.5pt}
\resizebox{0.45\textwidth}{!}{%
\begin{tabular}{l|ccc}
\hlineB{2}
\multicolumn{1}{c|}{System} & Accuracy(\%) & EER(\%) & C\textsubscript{avg} \\ \hlineB{2}
i-vector (CDS) & 60.32 & 26.98 & 26.35 \\
i-vector (GB) & 58.58 & 22.50 & 22.56 \\
i-vector-LDA (CDS)     & 62.60 & 21.05 & 20.12\\
i-vector-LDA (GB)     & 63.94 & 19.45 & 19.17\\
End-to-End (MFCC)    & 71.05 & 18.01 & 17.97 \\
\textbf{\begin{tabular}[c]{@{}l@{}}End-to-End (FBANK)\\ \end{tabular}}  & \textbf{73.39} & \textbf{16.30} & \textbf{15.96} \\
End-to-End (Spectrogram)    & 70.17 & 17.64 & 17.27 \\ \hlineB{2}
\end{tabular}%
}
\caption{Acoustic feature performance measurement on Arabic dialect MGB-3 dataset. All end-to-end systems were trained using random segmentation and volume/speed data augmentation.}
\label{tab:perform_all}
\end{table}

\begin{table}[t]
\centering
\setlength\doublerulesep{0.5pt}
\resizebox{0.45\textwidth}{!}{%
\begin{tabular}{c|l|ccc}
\hlineB{2}
\begin{tabular}[c]{@{}c@{}}Phoneme\\ Recognizer\end{tabular} & \multicolumn{1}{c|}{System} & Accuracy(\%) & EER(\%) & C\textsubscript{avg} \\ \hlineB{2}
\multirow{2}{*}{Hungarian} & Baseline & 48.86 & 29.94 & 29.16 \\ 
 & \textbf{Embedding} & \textbf{54.49} & \textbf{28.69} & \textbf{27.77} \\ \hline
\multirow{2}{*}{Russian} & Baseline & 45.04 & 31.30 & 30.65 \\
 & Embedding & 44.24 & 38.09 & 36.69 \\ \hline
\multirow{2}{*}{English} & Baseline & 33.78 & 39.46 & 39.04 \\
 & Embedding & 48.86 & 42.02 & 36.76 \\ \hline
\multirow{2}{*}{Czech} & Baseline & 45.64 & 31.85 & 31.16 \\ 
 & Embedding & 53.55 & 29.26 & 28.67 \\ \hlineB{2}
\end{tabular}%
}
\caption{Phoneme feature based on 4 different phoneme recognizer performance measurement.}
\label{tab:phoneme}
\end{table}

\begin{table}[t]
\centering
\resizebox{0.47\textwidth}{!}{%
\label{tab3}
\begin{tabular}{l|l|ccc}
\hlineB{2}
\multicolumn{1}{c|}{Feature} & \multicolumn{1}{c|}{System}               & Accuracy(\%) & EER(\%) & C\textsubscript{avg} \\ \hlineB{2}
\multirow{2}{*}{Character} & Baseline &51.34&30.03&30.17\\
                           & \textbf{Embedding}  &\textbf{58.18}&\textbf{25.48}&\textbf{25.68}\\ \cline{1-2} \hline
\multirow{2}{*}{Word}      & Baseline &50.00&30.73&30.41\\
                           & \textbf{Embedding}  &\textbf{58.51}&\textbf{24.87}&\textbf{24.99}\\ \hlineB{2}
\end{tabular}
}
\caption{Character and word feature performance measurement.}
\label{tab:char}
\end{table}

\begin{table}[t]
\centering
\setlength\doublerulesep{0.5pt}
\resizebox{0.47\textwidth}{!}{%
\begin{tabular}{c|ccc}
\hlineB{2}
\begin{tabular}[c]{@{}c@{}}Fusion system\\(\textbf{Bold} : end-to-end system \\ \textit{italic} : language embedding) \end{tabular}        & \multicolumn{1}{l}{Accuracy(\%)} & \multicolumn{1}{l}{EER(\%)} & \multicolumn{1}{l}{C\textsubscript{avg}} \\ \hlineB{2}
\textbf{FBANK} + \textit{word}& 76.94& 13.66 & 13.57 \\ 
\textbf{FBANK} + \textit{char}& 76.61& 13.89 & 13.87 \\ 
\textbf{FBANK} + \textit{phoneme}&75.13& 14.95& 14.79\\ 
\textbf{FBANK} + \textbf{MFCC}& 74.40 & 15.63 & 15.50\\ \hline
\textbf{MFCC} + \textit{word} + \textit{char} + \textit{phoneme} & 77.48& 14.02 & 14.00 \\
\textbf{FBANK} + \textit{word} + \textit{char} + \textit{phoneme} & \textbf{78.15}& \textbf{12.77} & \textbf{12.51} \\
\textbf{Spectrogram} + \textit{word} + \textit{char} + \textit{phoneme} & 77.88& 13.34 & 13.24 \\ \hline
i-vector + \textbf{FBANK} + \textit{word} + \textit{char} + \textit{phoneme} & \textbf{81.36}& \textbf{11.03} & \textbf{10.90} \\ \hlineB{2}
\end{tabular}%
}
\caption{Performance measurement of score fusion systems with end-to-end system and language embeddings.}
\label{tab:fusion}
\end{table}

\begin{table}[t]
\centering
\resizebox{0.4\textwidth}{!}{%
\begin{tabular}{l|cc}
\hlineB{2}
\multicolumn{1}{c|}{\multirow{2}{*}{Systems}} & \multicolumn{2}{c}{Accuracy(\%)} \\ \cline{2-3} 
\multicolumn{1}{c|}{} & Single System & Fusion System \\ \hlineB{2}
Khurana et al.~\cite{Khurana2017} & 67 & 73 \\
Shon et al.~\cite{Shon2017c} & 69.97 & 75.00 \\
Najafian et al.~\cite{Maryam2018} & 59.72 & 73.27 \\
Bulut et al.~\cite{Bulut2017} & - & 79.76 \\
Our approach & \textbf{73.39} & \textbf{81.36} \\ \hlineB{2}
\end{tabular}%
}
\caption{Performance comparison to previous works on same MGB-3 dataset}
\label{tab:compare}
\end{table}

All scores were normalized by Z-norm for each dialect, with fusion was done by logistic regression. Performance was measured in accuracy, Equal Error Rate (EER) and minimum decision cost function C\textsubscript{avg}. Accuracy was measured by taking the dialect showing the maximum score between a test utterance and the 5 dialects. Minimum C\textsubscript{avg} was computed from hard decision errors and a fixed set of costs and priors from~\cite{lre15}

\subsection{Experimental Result and Discussion}
For acoustic features, Figures \ref{fig:accuracy_feat} and \ref{fig:loss_feat} show the test set accuracy and mini-batch loss on training set for corresponding epochs. From the figures, it appears that the test set accuracy is not exactly correlated with loss and all three features show better performance before the network converged. We measured performance for two conditions denoted as "Maximum" and "Converged" in Table \ref{tab:perform_e2e}. The maximum condition means the network achieves the best accuracy on the validation set. The converged condition means the average loss of 100 mini-batches goes under 0.00001. From the table, the difference between maximum and converged is 5-10\% in all scenarios, so a validation set is essential to judge the performance of the network and should always be monitored to stop the training. 

MFCC and FBANK features achieve similar DID performance while the spectrogram is worse. Theoretically, spectrograms have more information than MFCCs or FBANKs, but it seems hard to optimize the network using the limited dataset. To obtain better performance on spectrograms, it seems to require using a pre-trained network on an external large dataset like VGG as in~\cite{Nagraniy2017}.

Figure \ref{fig:random_segment} shows performance gains from random segmentation. Random segmentation shows most gain when the test utterance is very short such as under 10 seconds. For over 10 seconds or original length, it still shows performance improvement because random segmentation provides diversity given a limited training dataset. However the impact is not as great as on short utterances.

Figure \ref{fig:accuracy_aug} and Table \ref{tab:perform_e2e_aug} show the performance on data augmented by speed and volume perturbation. The accuracy fluctuated similarly to that observed on the original data as in Figure \ref{fig:accuracy_feat}, but all numbers are higher. Volume perturbation shows slightly better results than using the original dataset in both the Maximum and Converged condition. However, speed perturbation shows impressive gains on performance, particularly, a 15\% improvement on EER. Using both volume and speed seem to have advantages when the network has converged, but it need more epochs to converge because of the larger data size compared to volume or speed perturbation.

Table \ref{tab:perform_e2e_aug_byfeat} shows a performance comparison on different features when using volume and speed data augmentation. The performance gains from dataset augmentation are 8\%, 11\% and 20\% in accuracy for MFCC, FBANK and spectrogram respectively. Spectrograms still attain the worst performance, as on the original dataset (table~\ref{tab:perform_e2e}). However it is somewhat surprising that the gain from increasing the dataset size is much higher than for MFCCs which achieve a relatively small increase.
Table \ref{tab:perform_all} shows performance comparisons between i-vectors and the end-to-end approach. A combination of i-vectors and LDA shows better performance on EER and C\textsubscript{avg} than the original i-vector. However, the end-to-end system in Table~\ref{tab:perform_e2e} (which uses the same original dataset as the i-vector system) outperforms the i-vector-LDA for all features. 

All end-to-end systems in Table~\ref{tab:perform_all} made use of random segmentation and volume/speed data augmentation during training. Since the internal dataset augmentation approach gives diversity on limited datasets, the end-to-end system could be significantly improved without any additional dataset. It is interesting that the most refined feature, MFCC, shows less performance improvement after dataset augmentation was applied (8\% in accuracy) than less refined, more raw features, i.e. FBANK and spectrogram show comparatively higher improvement by 14\% and 22\%.  Finally the spectrogram feature performs slightly better than MFCC in EER and C\textsubscript{avg}. It implies that if we have large dataset, we can use raw signals as input features. At the same time, however, it is difficult to determine how much training data is required for training raw features.

Tables \ref{tab:phoneme} and \ref{tab:char} show performance evaluations on language embeddings based on phoneme, character and word features. These features are also compared with the baseline system. Language embeddings show an average 17\% improvement on all metrics. Language embeddings based on word features achieve the best performance among the three features. Another benefit is that the linguistic feature dimension can be significantly reduced. For example, the 41,657 dimension character feature was reduced to 200 and is only 0.5\% of the original dimensionality, though the EER improves about 15\%.

Table \ref{tab:fusion} shows the performance of fusion systems with various combinations. We use Hungarian phonemes for phonetic-based language embeddings which attains the best performance among phoneme features. The end-to-end system used augmented data as shown in Table~\ref{tab:perform_all}. From the result, it is clear that fusion between acoustic and language embeddings shows better efficiency than fusion between end-to-end systems such as MFCC and FBANK, although the performance on language embeddings itself shows comparably lower performance than acoustic end-to-end systems. Finally, all the scores from each feature were fused, with the fusion system using FBANK features achieving better performance than MFCC or spectrogram-based fusion systems on all measurements. Also, we find that the spectrogram system performs slightly better than the MFCC counterpart. 

Although fusion of multiple systems is somewhat not feasible for a practical situation, we fused 5 different systems (i-vector with LDA and GB, FBANK based end-to-end system, language embeddings) and verified the best performance compared to previous studies. Table~\ref{tab:compare} shows summary performance of previous studies on same MGB-3 Arabic dialect dataset. Both single and fusion system show outstanding performance than others.  

\section{Conclusion}

In this paper, we describe an end-to-end dialect identification system using acoustic features and language embeddings based on text-based linguistic features. We investigated several acoustic and linguistic features along with various dataset augmentation techniques using a limited dataset resource. We verified that the end-to-end system based on acoustic feature outperforms i-vectors and also that language embeddings derived from a Siamese neural network boost the performance by learning the similarities between utterances with the same dialect and dissimilarities between different dialects.  Experiments on the MGB-3 Dialectical Arabic corpus show that the best single system achieves 73\% accuracy, and the best fusion system shows 78\% accuracy. From the experiments, spectrograms could be utilized as acoustic feature when the training dataset is large enough.  We also observe that the end-to-end dialect identification system can be significantly improved using random segmentation and volume/speed perturbation to increase the diversity and amount of training data.  The end-to-end dialect identification system has a simplified topology and training methodology compared to a bottleneck feature based i-vector extraction scheme. Finally, using a Siamese network to learn language embeddings reduces the linguistic-feature dimensionality significantly, and provide synergistic fusion with acoustic features.

\bibliographystyle{IEEEbib}
\bibliography{mendeley.bib}

\end{document}